\documentclass[aps,prb,preprint,superscriptaddress,showpacs,showkeys]{revtex4}

\usepackage{graphicx}
\usepackage{bm}
\usepackage{amsmath}
\usepackage{amssymb}
\usepackage{hyperref}

\begin{document}

\title{Effects of substrate and electric fields on charges in nanotubes}

\author{Zhao Wang}
\email{zhao.wang@empa.ch,wzzhao@yahoo.fr}
\affiliation{Institute UTINAM, Universit\'{e} de Franche-Comt\'{e}, Besan\c{c}on, France.}
\affiliation{EMPA (Swiss Federal Laboratories for Materials Testing and Research), Thun, Switzerland}

\begin{abstract}
In this paper, we study how the distribution of net charges in carbon nanotubes can be influenced by substrate and external electric fields, using theoretical calculations based on an extension of the atomic charge-dipole model. We find that the charge enhancement becomes less significant when the tube gets closer to substrate or when the dielectric constant of substrate increases. It is demonstrated that net charges can be shifted to one side of tube by longitudinal electric fields and the polarity of charges can be locally changed, while transversal fields give much less influence on the charge enhancement. These properties could be generalized for other metallic or semiconducting nano/microwires and tubes. 
\end{abstract}

\pacs{73.63.Fg, 68.37.Ps, 85.35.Kt, 41.20.Cv}
\keywords{nanotubes, charges, substrate, electric fields}

\maketitle

\section{Introduction}
\label{intro}

The distribution of electric charges in carbon nanotubes (CNTs) is of interest for their future uses in nanoelectromechanical systems (NEMS)\cite{Anantram2006} such as field emission devices,\cite{De1995,purcell-02} sensors, \cite{Snow2005} actuators\cite{Roth2002} and charge storages.\cite{akita-01, Cui2002, jang-05, Lu2006, Ryu2007} Recently, electric force microscopies (EFM) have been used to inject and to detect net charges in CNTs,\cite{Paillet2005, Zdrojek2005} electric charges are found to distribute uniformly along CNTs. However, charge accumulation (so-called charge enhancement) at tube ends has been predicted by theoretical studies using density functional theory\cite{Keblinski2002} and classical electrostatics\cite{Li2006a} calculations. These predicted properties have been then confirmed by Zdrojek \textit{et al.}\cite{Zdrojek2006} in EFM experiments. It was also shown that electric charges can be trapped in CNT loops during periods of time\cite{Jespersen2005}. Furthermore, charge-induced failures\cite{Li2007} and structure changes of CNTs\cite{Gartstein2002} were reported. In one of our previous works, weak charge enhancement at the tube ends and its geometry dependence were demonstrated by the combination of theoretical calculations and EFM experiments.\cite{Wang2008}

In this paper, we address the issue of the substrate- and electric-field-effects on the charge distribution in CNTs, since CNTs are usually deposited on substrate and driven by electric fields in a number of nanodevices.\cite{Tans1998,Rueckes2000} It is known that substrate can exert quite strong influence on the charge distribution, as discussed recently in Ref.\cite{Mowbray2006} for the case of ions inside CNTs. Theoretical calculations have been performed due to the difficulties for accurately quantifying this effect in recent experiments. Our calculation results reveal that the charge enhancement becomes less significant when substrate gets closer to CNTs, and that the enhancement ratio decreases with increasing dielectric constant of substrate. These effects on the charge distribution in radical directions are also discussed. Furthermore, we find that the charge distribution in CNTs can be significantly modified in external fields. The dependence of field strength is demonstrated for both single-walled and multi-walled CNTs (SWCNTs and MWCNTs). We note that the properties demonstrated in this paper could also apply to semiconducting CNTs, because semiconducting and metallic nanotubes are both expected to accept extra charges, from theoretical\cite{Margine2006} and experimental\cite{Paillet2005, Zdrojek2006} points of view. 

The charge distribution has been computed using a Gaussian-regularized atomic charge-dipole interacting model.\cite{mayer-07-01, Mayer2008} It has been developed from the atomic dipole theory of Applequist\cite{Applequist-72} and has recently been parameterized for CNTs.\cite{mayer-05-01} In this model, each atom is treated as an interacting polarizable point with a free charge, the static equilibrium state of charges are determined by minimizing the total electrostatic energy of system. In this work, we have extended this model to take the substrate effect into account by including surface-induced terms to vacuum electrostatic interacting tensors using the method of mirror image.\cite{Jacksonbook1975} Compared with classical Coulomb-law-based models in which only the charge is considered, this model provides a more accurate description of electrostatic properties of CNTs, because not only the net charges, but also the induced dipole, atomic polarizabilities and the image charge are taken into account. 

For the outline, our computational model is presented in sec.\,II. Results for the effects of substrate and fields are discussed in sec.\,III and sec.\,IV, respectively. We draw a conclusion in sec.\,V. The formulation of the surface-induced electrostatic interacting tensors is given in Appendix.

\section{Computational model}
\begin{figure}[htp]
\centerline{\includegraphics[width=10cm]{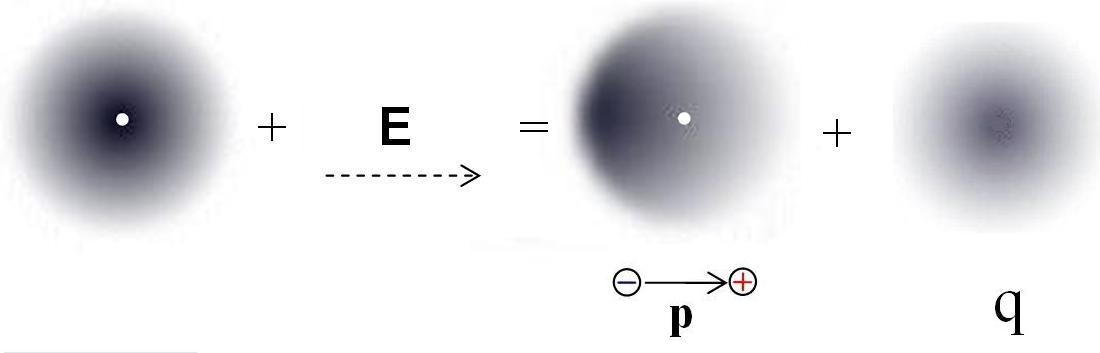}}
\caption{\label{fig:chargedipoles}
Schematic of the principle of the charge-dipole model, in which each atom is modeled as a net charge $q$ with an induced dipole $\bm{p}$.
}
\end{figure}

In our calculation, each atom is associated with an electric charge $q$ and an induced dipole $\bm{p}$ as shown in Fig. \ref{fig:chargedipoles}. 

The total electrostatic energy $U^{elec}$ for a CNT of $N$ atoms can be written as follows:

\begin{multline}
\label{eq:1}
U^{elec}=\sum_{i=1}^N{q_i(\chi_i+V_i)}-\sum_{i=1}^N{\bm{p}_i\cdot\bm{E_i}}+
\frac{1}{2}\sum_{i=1}^N{\sum_{\substack{j=1}}^N{q_i T^{i,j}_{q-q} q_j}}\\-\sum_{i=1}^N{\sum_{\substack{j=1}}^N{\bm{p}_i \cdot \bm{T}^{i,j}_{p-q} q_j}}
-\frac{1}{2}\sum_{i=1}^N{\sum_{\substack{j=1}}^N{\bm{p}_i \cdot \bm{T}^{i,j}_{p-p} \cdot \bm{p}_j}}
\end{multline}

where $\chi$ is the electron affinity, $V$ and $\bm{E}$ stand for the external potential and electric field, respectively. $T$ and $\bm{T}$ are the electrostatic interacting tensors. They can be written as
$T^{i,j}_{q-q}=(1 / 4 \pi \varepsilon_{0}) \times (1/r_{ij})$,  $\bm{T}^{i,j}_{p-q}=-\nabla_{\bm{r}_i} T^{i,j}_{q-q}$ and $\bm{T}^{i,j}_{p-p}=-\nabla_{\bm{r}_j} \otimes \nabla_{\bm{r}_i} T^{i,j}_{q-q}$, where $r_{i,j}=\left| \bm{r}_{j}-\bm{r}_{i} \right|$. We have regularized $T$ and $\bm{T}$ by a Gaussian distribution in order to avoid divergence problems when atoms are too close to each other, as discussed previously in Refs.\cite{mayer-07-01,Wang2007a}. Note that the value of Gaussian charge distribution width $R$ used in this work for free-end atoms is fitted to $0.1273$ nm (about 1.3 time that of the carbon atom with three chemical bonds) from results in a previous study using DFT calculation.\cite{Keblinski2002}

The equilibrium state of charges and dipoles should correspond to the minimum value of $U^{elec}$, and hence the derivatives of $U^{elec}$ with respect to $q$ and $\bm{p}$ should be zero. Taking this boundary condition as well as total molecular net charge $Q^{tot}$ into account with the self-energy terms (when $i=j$), we can obtain the equilibrium configuration of charge and dipole by solving $N$ linear vectorial equations and $N+1$ linear scalar equations as follows: 

\begin{equation}
\label{eq:2}
\begin{array}{c}
\left\{ \begin{array}{ll}
\sum\limits_{j=1}^N{\bm{T}^{i,j}_{p-p}\otimes\bm{p}_{j}}+\sum\limits_{j=1}^N{\bm{T}^{i,j}_{p-q}q_{j}}=-\bm{E_i}\\
\sum\limits_{j=1}^N{\bm{T}^{i,j}_{p-q}\cdot\bm{p}_{j}}+\sum\limits_{j=1}^N{T^{i,j}_{q-q}q_{j}}+\lambda = -(\chi_i+V_i)\\
\sum\limits_{j=1}^N{q_{j}}=Q^{tot}
\end{array}\right.\\
\forall{i=1,...,N}
\end{array}
\end{equation}

where $\lambda$ is a Lagrange multiplier,\cite{Lagrange1797} which is related to the chemical potential of the molecule. In case of a CNT close to a substrate, as in EFM experiments,\cite{Zdrojek2005,Paillet2005,Jespersen2005} the distributions of charges and dipoles are different from those in free space. We have taken this boundary condition into account by adding a surface-induced terms $T^{m}$ and $\bm{T}^{m}$ to the vacuum electrostatic interacting tensors using the method of mirror images.\cite{Jacksonbook1975} The detailed formulation of $T^{m}$ and $\bm{T}^{m}$ can be found in Appendix. Note that the substrate surface is assumed to be infinitely plane in this work. 

The structures of CNTs are relaxed by means of energy optimization\cite{Wang2007a} using the conjugated gradient method\cite{Payne1992} based on the AIREBO (adaptive interatomic reactive empirical bond order) potential\cite{stuart-00}.

\section{Influence of substrate}

\begin{figure}[htp]
\centerline{\includegraphics[width=12cm]{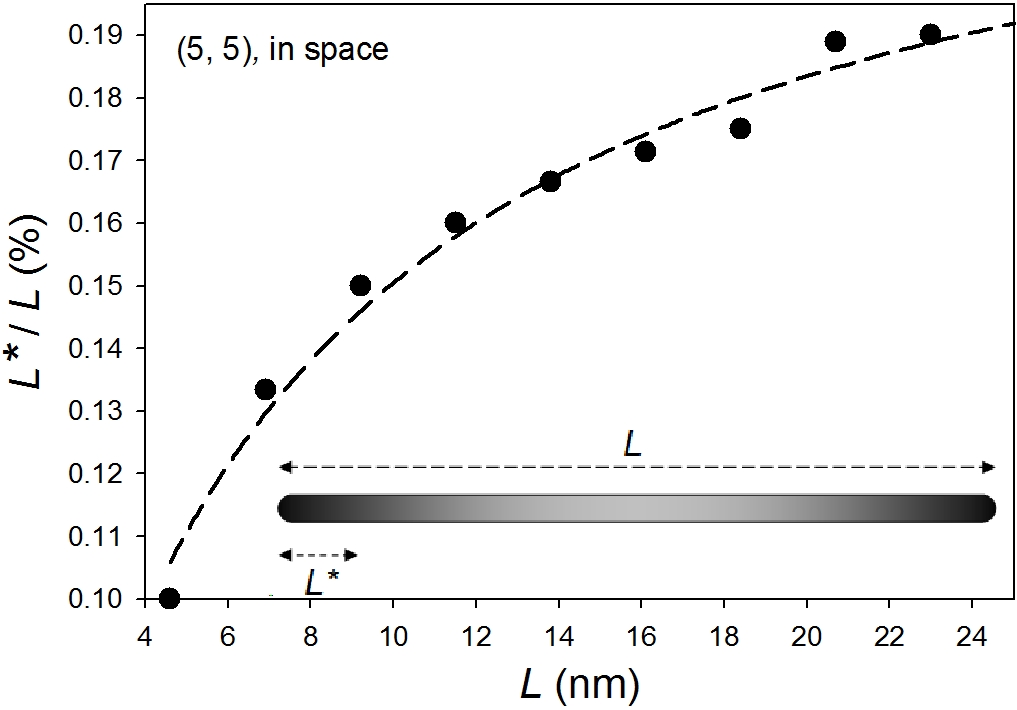}}
\caption{\label{fig:Lenhance}
$L$ stands for the tube length and $L^{*}$ is the length of the enhancement zone in which the average charge density is higher than the average $\sigma^{ave}$ over the whole tube. The circles present the calculated points. Inset: net charge density along a freestanding tube in space.
}
\end{figure}

Previous studies show static charge accumulations at tube ends\cite{Keblinski2002, Zdrojek2008} (as shown in the insert of Fig.\,\ref{fig:Lenhance}). A well-defined zone of the charge accumulation is required in order to well quantify this enhancement effect. In Fig.\,\ref{fig:Lenhance}, we can see that the length of charge enhancement zone ($L^{*}$) increases with the tube length ($L$), if we define this zone as the part where the charge density is higher than the average over the whole tube ($\sigma^{ave}$). 

Considering that the CNTs used in experiments are usually longer than those used in our calculation, we define the length of charge enhancement zone as 20\% of $L$ (10\%$L$ at each tube end). The ratio of charge enhancement\cite{Wang2008} is denoted as follows:

\begin{equation}
\label{eq:3}
\varphi=\sigma^{end}/\sigma^{middle}
\end{equation}

where $\sigma^{end}$ is the average charge density in the enhancement zone (10\%$L$ at each tube end), and $\sigma^{middle}$ is that at the middle of the tube. We note that $\varphi$ is independent of $\sigma^{ave}$, because the local charge densities are proportional to $\sigma^{ave}$ with respect to a constant electric potential on the tube surface.

\begin{figure}[htp]
\centerline{\includegraphics[width=12cm]{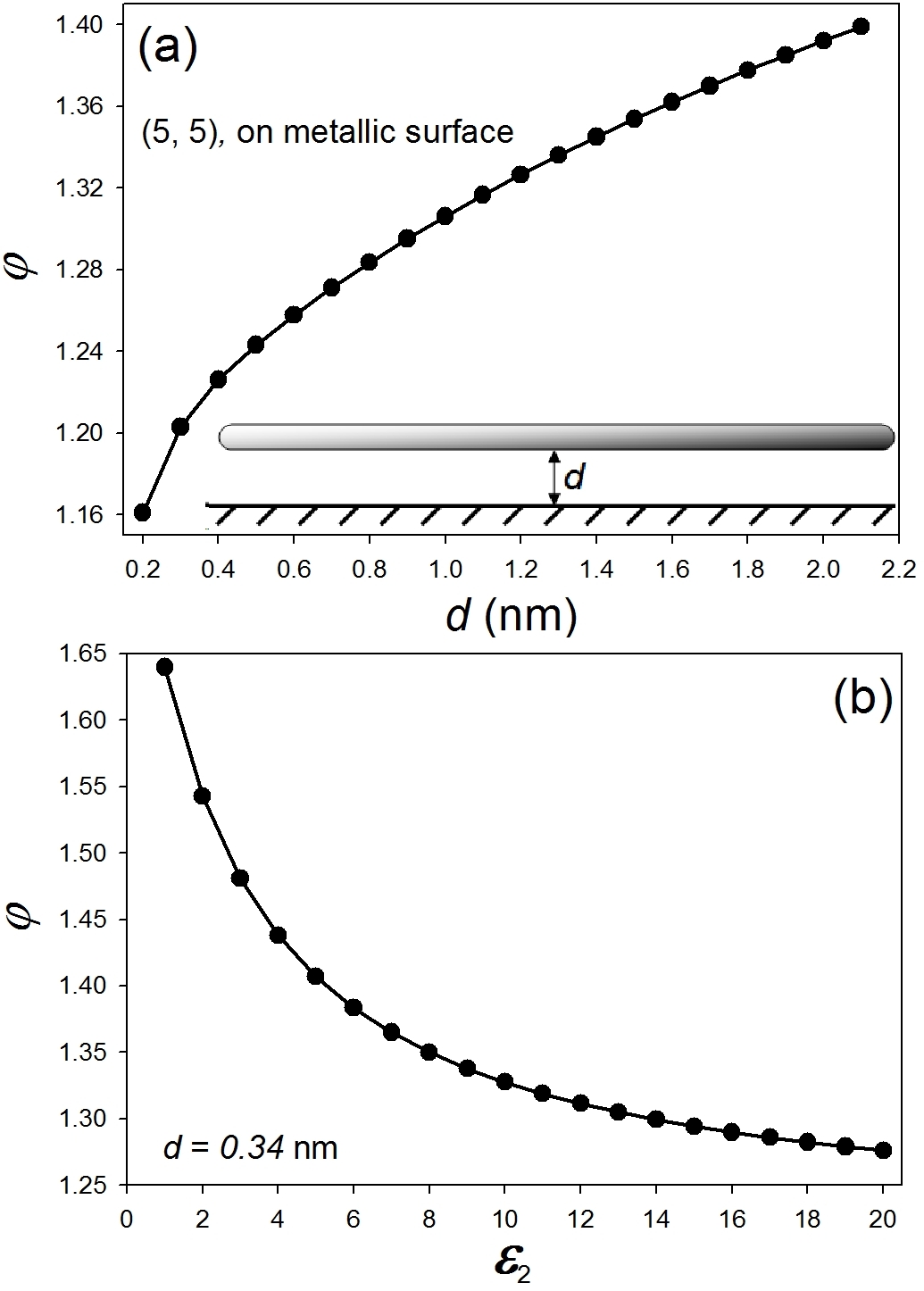}}
\caption{\label{fig:Surface_dep}
(a) $\varphi$ \textit{versus} $d$, for a charged (5, 5) SWCNT ($L\approx27$ nm) with $\sigma^{ave}=0.55 \times 10^{-3} e$/atom and $\varepsilon_{2}=\infty$. Inset: net charge density in a nanotube in a semi-infinite space. (b) $\varphi$ \textit{versus} $\varepsilon_{2}$, for the same tube ($d=0.34$nm).
}
\end{figure}

In case of a CNT in a semi-infinite space (e.g. deposited on substrate), net charges will be attracted to the tube bottom by opposite image charges appearing on substrate surface, as shown in the inset of Fig. \ref{fig:Surface_dep} (a). This surface effect mainly depends on the tube-surface physisorption distance\cite{Rafii-Tabar2004, Tsetseris2006} ($d$) and the dielectric constant of the substrate\cite{Li2006a, Besteman2005} ($\varepsilon_{2}$). Both of them vary with the type of substrate material. To demonstrate the influence of these two parameters on charge enhancement, we plot $\varphi$ \textit{versus} $d$ and $\varepsilon_{2}$ in Fig. \ref{fig:Surface_dep} (a) and (b), respectively. We can see in Fig. \ref{fig:Surface_dep} (a) that the charge enhancement becomes less significant when the substrate surface gets closer to the tube (when $d$ decreases). From electrostatic point of view, the main mechanism of this effect is that the charge distributed area (band) in radical direction has been effectively reduced since a part of net charges is attracted to the tube bottom, and hence the charge distribution along the tube axis gets closer to that along an infinite-long tube, in which the charge distribution is perfectly uniform ($\varphi=1$). Similar behavior can be contrasted with the situation when $\varepsilon_{2}$ increases, as shown by the plot in Fig.\,\ref{fig:Surface_dep} (b). This implies that $\varphi$ can get higher if one uses small-dielectric constant material instead of metal ($\varepsilon_{2}=\infty$) in experiments, and that the CNT exhibits the strongest charge enhancement in an infinite space ($\varepsilon_{2}=0$).    
 
\begin{figure}[htp]
\centerline{\includegraphics[width=12cm]{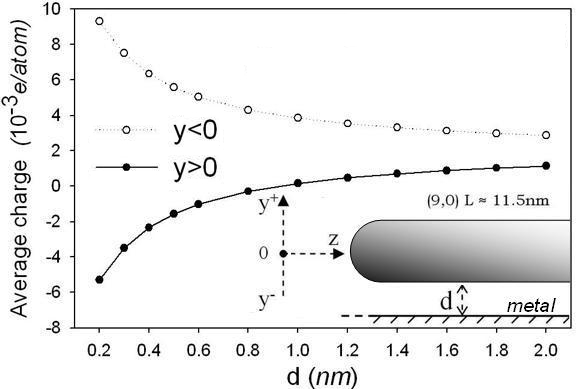}}
\caption{\label{fig:surface4}
Average charge density \textit{versus} $d$ for $y>0$ (the top half part) and $y<0$ (the bottom half part), respectively, for a close-ended (9,0) SWCNT ($L\approx12$ nm $\sigma^{ave}\approx2.0 \times 10^{-3} e$/atom) on a metallic surface. Inset: schematic of transversal charge distribution at a tube end.
}
\end{figure}

The issue about the charge distribution in radical(transversal) direction has rarely been discussed in the literature, although it is one of the main mechanisms of discharging phenomenon observed experimentally.\cite{Zdrojek2005, Paillet2005} When a CNT is horizontally deposited upon a substrate, charges migration is mainly caused in the direction perpendicular to the substrate surface, as a typical mirror effect (see Fig.\,\ref{fig:surface4}). Local charge accumulation at the bottom of tube ends (open circles) can directly lead to enhanced electron emission.\cite{Charlier2002} We can also see that the top part of the tube ($y>0$) even shows opposite electric sign when $d\leq0.8$ nm (solid circles). 

\begin{figure}[htp]
\centerline{\includegraphics[width=15cm]{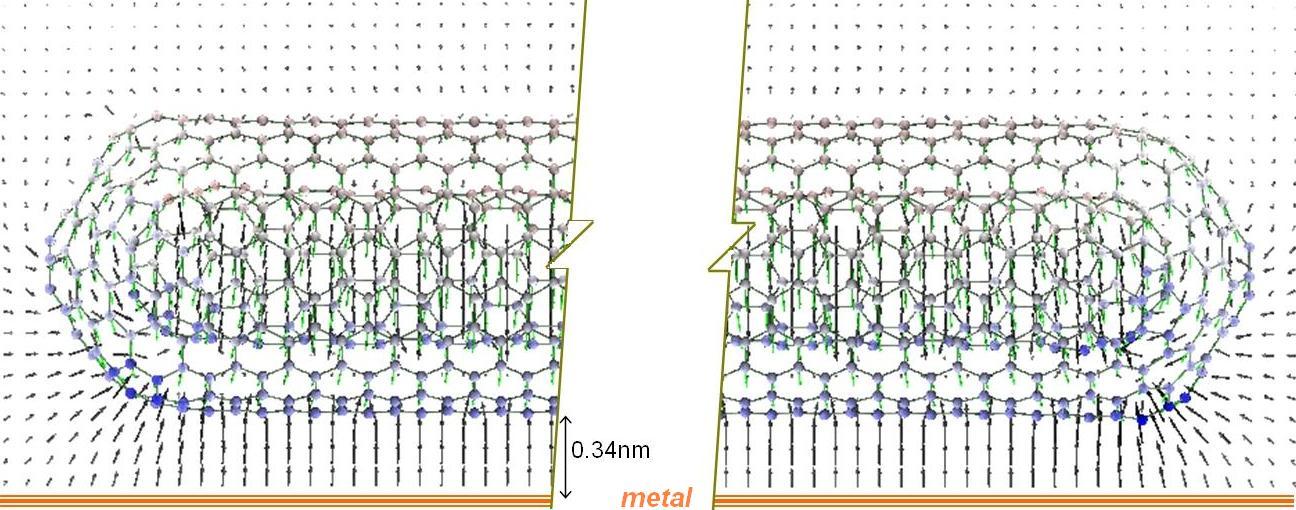}}
\caption{\label{fig:SurfaceDWCNT}
(Color online) Atomic graph of the charge distribution in a (9,0)@(10,10) DWCNT ($L\approx10.3$ nm, $\sigma^{ave}=0.55\times 10^{-3} e$/atom for both two layers) on a metallic surface. Color of the atoms is proportional to the local charge density. Dark arrows stand for the local electric fields induced by the net charge around the tube ends, their length and color are proportional to the field intensities. The maximum atomic charge densities are: $\sigma^{max}=8 \times 10^{-3}e$/atom. The maximum strength of the local electric fields presented in this figure  $E^{max}=4.3$ V/nm. 
}
\end{figure}

The issue about charge distribution in MWCNTs is more complicated due to depolarization, field screening and electrostatic interactions between layers. To show further details about substrate effects, we depict the atomic charge distribution in a double-walled CNT (DWCNT) electrically charged in its both inner and outer carbon layers in Fig.\,\ref{fig:SurfaceDWCNT}. The migration of atomic charges induced by the metallic surface is shown in this figure. We can see the enhanced local electric fields around the tube bottom due to the charge enhancement. The top of the tube even shows electrically positive since most of net charges (negative) are attracted to the tube bottom by the surface images.\cite{Lang1973} 

\section{Influence of external fields}
Recent works showed that external electric fields could induce alignments\cite{joselevich-02}, deformations\cite{Poncharal-99}, field emission\cite{Rinzler1995} and conductivity transitions\cite{Rochefort2001} of CNTs. Here we concern mainly on how electric fields influence the static distribution of net charges in CNTs. 

\begin{figure}[htp]
\centerline{\includegraphics[width=15cm]{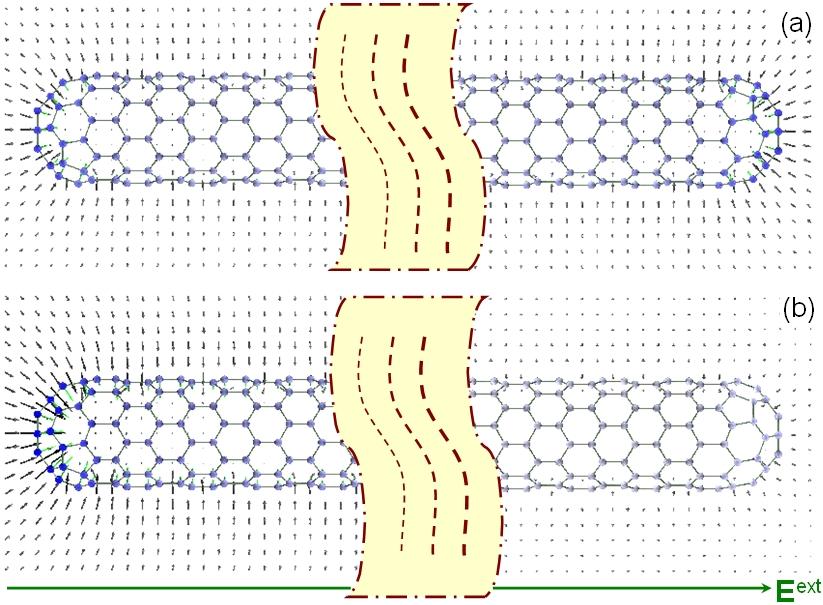}}
\caption{\label{fig:mapfield}
(Color online) Charge distribution in a charged (9,0) SWCNT ($L\approx12$ nm, $\sigma^{ave}=1.0\times 10^{-3} e$/atom), (a) in space, and (b) in an axial uniform external electric field $E^{ext}=0.05$ V/nm. Color of atoms is proportional to charge density. Dark arrows stand for local electric fields around tube ends, their length and color are proportional to field intensity.}
\end{figure}

The charge distribution of a SWCNT in free space is compared with that in an external electric field $\bm{E}^{ext}$ in Fig.\,\ref{fig:mapfield}. As expected, net charges are shifted to one side, around which local electric fields (fields induced by net charges and dipoles $+$ external fields) are enhanced. The magnitude of this polarization effect is roughly proportional to the external field intensity $E$ and the tube length $L$,\cite{benedict-95} this implies that, for CNTs used in experiments (usually $L \thicksim \mu$m), $\bm{E}^{ext}$ can be hundreds times weaker for producing a similar effects as those shown in Fig.\,\ref{fig:mapfield}. Moreover, we note that $\bm{E}^{ext}$ used in this work is about two magnitudes weaker than that can lead to field emission from our short CNTs.\cite{Jo2003}

\begin{figure}[htp]
\centerline{\includegraphics[width=12cm]{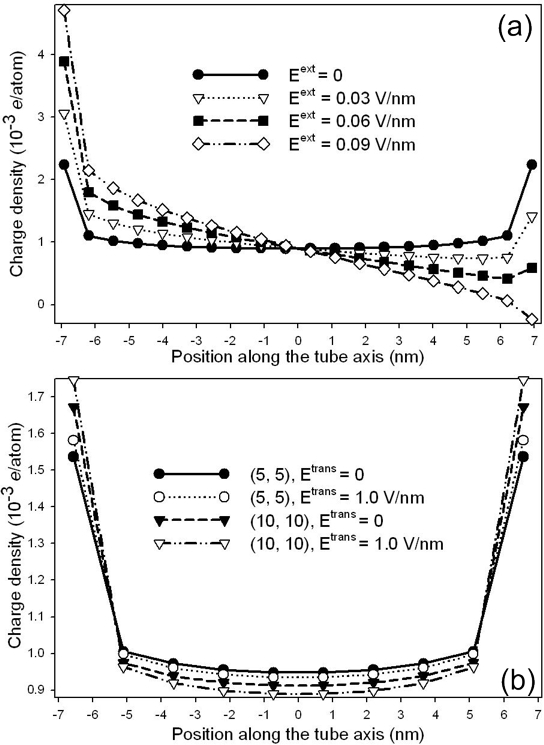}}
\caption{\label{fig:axialfield}
Charge profile along a charged (5,5) SWCNT ($L \approx 14$ nm, $\sigma^{ave}=0.9\times 10^{-3} e/$atom). (a) In longitudinal (along the tube axis) electric fields $\bm{E}^{ext}$. Each point is calculated as the average of 5\%$L$. (b) In transversal (perpendicular to the tube axis) electric fields $E^{trans}$. Each point is calculated as the average of 10\%$L$. 
}
\end{figure}

To achieve a quantitative comparison, we plot in Fig.\,\ref{fig:axialfield} charge distribution along a SWCNT in an external field $\bm{E}^{ext}$. It can be seen that the typical \textit{U}-like distribution\cite{Keblinski2002} in \textit{vacuum} (solid circles) can be significantly modified by the axial external electric field. On the other hand, the influence of transversal electric fields is expected to be weak, due to the strong anisotropy of the polarizabilities of CNTs. We can see in Fig.\,\ref{fig:axialfield} (b) that, even with very strong field intensities in the order of V/nm, the charge profile does not change a lot. In fact, the transversal field mainly influences the charge distribution in non-axial direction. Moreover, it needs to mention that the average charge density depends on the value of unit length taken in the calculation, e.g. the value of charge density represented by the solid circles in Fig.\,\ref{fig:axialfield} (b) is lower than that in Fig.\,\ref{fig:axialfield} (a),  because it is calculated as average on every $10$\%$L$, instead of that on every $5$\%$L$.

\begin{figure}[htp]
\centerline{\includegraphics[width=12cm]{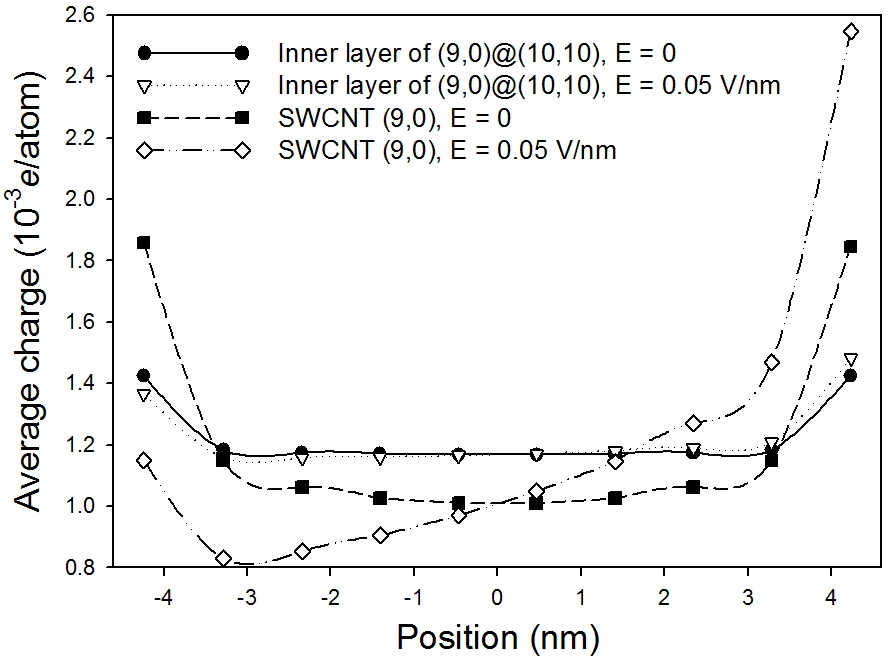}}
\caption{\label{fig:MWCNT}
\textit{top.} Charge profile along the tube axis, for the inner tube of a DWCNT (9,0)@(10,10) and for a SWCNT (9,0) with the same size. 
}
\end{figure}

For MWCNTs, it has been reported that electric screening plays an important role for the field effects.\cite{kozinsky-06} To demonstrate its influence on the charge enhancement, we compare the charge distribution in an inner layer of a DWCNT with that of a SWCNT with the same size in Fig.\,\ref{fig:MWCNT}. In this comparison, it can be seen that the effect of external fields is much weaker on the charge distribution in the inner layer of the DWCNT, due to electrostatic screening. Furthermore, by comparing the longitudinal charge distribution with the SWCNT, we have found that the charge enhancement is lower in the DWCNT, due to the electric repulsive interaction between the two carbon layers. It is a typical `` $1+1<2$ '' effect.\cite{Marinopoulos2003,Pfeiffer2004}

\section{Conclusion}
In summery, influences of substrate and external electric fields on electric charges in CNTs have been investigated by extending the charge-dipole polarization model. The results obtained are relevant for a better understanding of the distribution and stability of electric charges in CNTs in possible experimental situations (e.g. CNTs deposited on a solid surface). Local charge enhancement at tube ends is studied as a particular effect. Our results reveal that the charge enhancement becomes less significant when the substrate-tube separation decreases or when the dielectric constant of substrate increases. Charge delocalization in radical direction has been observed as a typical mirror effect in presence of substrate or transversal electric fields. Longitudinal external electric fields have been found to have much more influence on the charge enhancement than the transversal ones with same intensities. Electric screening in MWCNTs is found to influence charge profile in MWCNTs, especially in presence of electric fields. In general, these above conclusions could also qualitatively apply to other nanowires and tubes, from electrostatic point of view. 

\section{acknowledgments}
Dr. M. Devel, Dr. M. Zdrojek, Dr. T. M\'{e}lin and Dr. A. Mayer are gratefully acknowledged for useful discussion.

\section{Appendix: Surface-induced terms of electrostatic interaction tensors}
\begin{figure}[htp]
\centerline{\includegraphics[width=10cm]{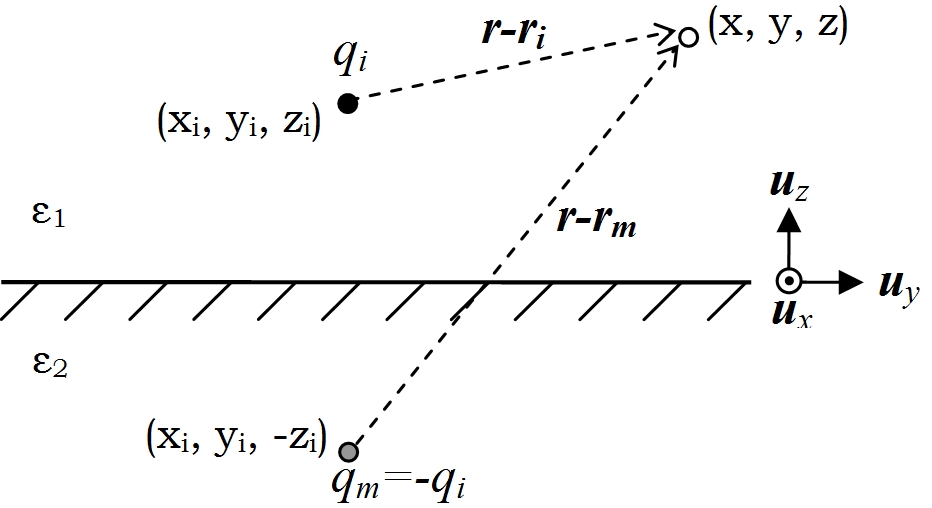}}
\caption{\label{fig:appendix}
Schematic of the charge image.
}
\end{figure}

To take substrate effects into account, we have extended the charge-dipole model of Mayer\cite{mayer-07-01} by adding surface-induced terms (${T}_{0}^{m}$, $\bm{T}^{m}_{1}$ and $\bm{T}^{m}_{2}$) into the vacuum electrostatic interaction tensors (${T}_{0}$, $\bm{T}_{1}$ and $\bm{T}_{2}$), respectively, using the method of mirror image.\cite{Jacksonbook1975} In this method, the electric potential $V^{m}$ on an arbitrary point ($x,y,z$) induced by the mirror image of a point charge $q_{i}$ embedded in a semi-infinite medium $\varepsilon_{1}$ close to another medium $\varepsilon_{2}$ (see Fig.\,\ref{fig:appendix}) can be written as follows:

\begin{equation}
\label{eq:5}
V^{m}(\bm{r})=-k q_{i}\frac{1}{\left|\bm{r}-\bm{r}_{m}\right|} = q_{i}{T}_{0}^{m}
\end{equation}

where $\bm{r}_{m}$ is the coordinate of the mirror image $q_{i}^{m}$ and $k = (\varepsilon_{2} - \varepsilon_{1}) / \left[4\pi \varepsilon_{0}(\varepsilon_{2} + \varepsilon_{1}) \right]$ is an electrostatic constant. ${T}_{0}^{m}$ stands for the $0$th order interaction tensor for the mirror image. It is the Green's function for the vectorial variable Laplace equation.
 
For our system shown in this Fig. \ref{fig:appendix} using Cartesian coordinate, the \textit{0th} order mirror-charge interaction tensor can be written as:

\begin{equation}
\label{eq:6}
{T}_{0}^{m}=\frac{-k}{\left|\bm{r}-\bm{r}_{m}\right|}=\frac{-k}{\sqrt{\delta x^{2}+\delta y^{2}+\zeta^{2}}}
\end{equation}

where $\delta x = x-x_{i}$, $\delta y = y - y_{i}$, $\zeta = z+z_{i}$. 

The interaction tensors of the \textit{1st} (charge-dipole) and the \textit{2nd} (dipole-dipole) order for the image charges can be derived from that of the \textit{$0$th} order (charge-charge):

\begin{equation}
\label{eq:7}
\bm{T}^{m}_{1}=-\nabla_{r_{m}}{T}_{0}^{m}=
\frac{-k}{\left|\bm{r}-\bm{r}_{m}\right|^{3}}
\left[
\begin{matrix}
\delta x\\
\delta y\\
-\zeta \\
\end{matrix}
\right]_{\bm{u}_{x},\bm{u}_{y},\bm{u}_{z}}
\end{equation}
and
\normalsize
\begin{equation}
\begin{array}{l}
\bm{T}^{m}_{2}= -\nabla_{r}\otimes\nabla_{r_{m}}{T}_{0}^{m}= \frac{k}{\left|\bm{r}-\bm{r}_{m}\right|^{5}}
\times \\
\left[
\begin{matrix}
\delta y^{2}+\zeta^{2}-2\delta x^{2} & -3\delta x\delta y & 3\delta x\zeta\\
-3\delta x\delta y & \delta x^{2}+\zeta^{2}-2\delta y^{2} & 3\delta y\zeta\\
-3\delta x\zeta & -3\delta y\zeta & 2\zeta^{2}-\delta x^{2}-\delta y^{2}\\
\end{matrix}
\right]
\end{array}
\end{equation}

We note that the vacuum interaction tensors used in present study are regularized by a normal distribution, in order to avoid divergence problems with point charges when atoms get too close to each other.\cite{mayer-07-01} However, it is not necessary to regularize the surface-induced terms since the distance between the net charge and its images is generally large enough.

\end{document}